%% file: ms.tex
\documentclass[useAMS,usenatbib]{mn2e}
\pdfoutput=0

\usepackage{graphicx}
\usepackage{aasmacros}
\usepackage{color}
\usepackage{epsfig}

\input{newcommands.tex}

\def\fq{f_{\rm Q}}
\def\fqcen{f_{\rm Q}^{\rm cen}}
\def\mgal{M_\ast}
\def\hhmsol{M_\odot/h^2}
\def\dn{{\rm D}_n4000}

\def\mhalo{M_{h}}
\def\lesssim{\la}

\def\zhalf{z_{1/2}}
\def\mpeak{M_{\rm peak}}
\def\cvir{c_{\rm vir}}

\title[Halo and Galaxy Histories I]{Halo Histories vs. Galaxy Properties
  at $z=0$\\ I: The Quenching of Star Formation}

\author[Tinker et.~al.]{Jeremy L. Tinker$^1$,
  Andrew
  R. Wetzel$^{2,3,4}$, Charlie Conroy$^5$, Yao-Yuan Mao$^{6}$\\
  $^1$Center for Cosmology and Particle Physics, Department of Physics, New York University, New York, NY\\
  $^2$TAPIR, California Institute of Technology, Pasadena, CA\\
 $^3$Carnegie Observatories, Pasadena, CA\\
$^4$Department of Physics, University of California, Davis, CA\\
  $^5$Department of Astronomy, Harvard University, Cambridge, MA\\
  $^6$Department of Physics and Astronomy \& Pittsburgh Particle Physics, Astrophysics, and Cosmology Center (PITT PACC),\\ University of Pittsburgh, Pittsburgh, PA 15260, USA}
\begin{document}


\pagerange{\pageref{firstpage}--\pageref{lastpage}} \pubyear{2016}

\maketitle

\label{firstpage}

\begin{abstract}

  We test whether halo age and galaxy age are correlated at fixed
  halo and galaxy mass. The formation histories, and thus ages, of
  dark matter halos correlate with their large-scale density $\rho$,
  an effect known as assembly bias. We test whether this correlation
  extends to galaxies by measuring the dependence of galaxy stellar
  age on $\rho$. To clarify the comparison between theory and
  observation, and to remove the strong environmental effects on
  satellites, we use galaxy group catalogs to identify central
  galaxies and measure their quenched fraction, $\fq$, as a function
  of large-scale environment. Models that match halo age to central
  galaxy age predict a strong positive correlation between $\fq$ and
  $\rho$. However, we show that the amplitude of this effect depends
  on the definition of halo age: assembly bias is significantly
  reduced when removing the effects of splashback halos---those halos
  that are central but have passed through a larger halo or
  experienced strong tidal encounters. Defining age using halo mass at
  its peak value rather than current mass removes these effects. In
  SDSS data, at $\mgal\ga 10^{10}$ $\hhmsol$, there is a $\sim 5\%$
  increase in $\fq$ from low to high densities, which is in agreement
  with predictions of dark matter halos using peak halo mass. At lower
  stellar mass there is little to no correlation of $\fq$ with
  $\rho$. For these galaxies, age-matching is inconsistent with the
  data across the wide range the halo formation metrics that we
  tested. This implies that halo formation history has a small but
  statistically significant impact on quenching of star formation at
  high masses, while the quenching process in low-mass central
  galaxies is uncorrelated with halo formation history.

\end{abstract}


\begin{figure*}
\psfig{file=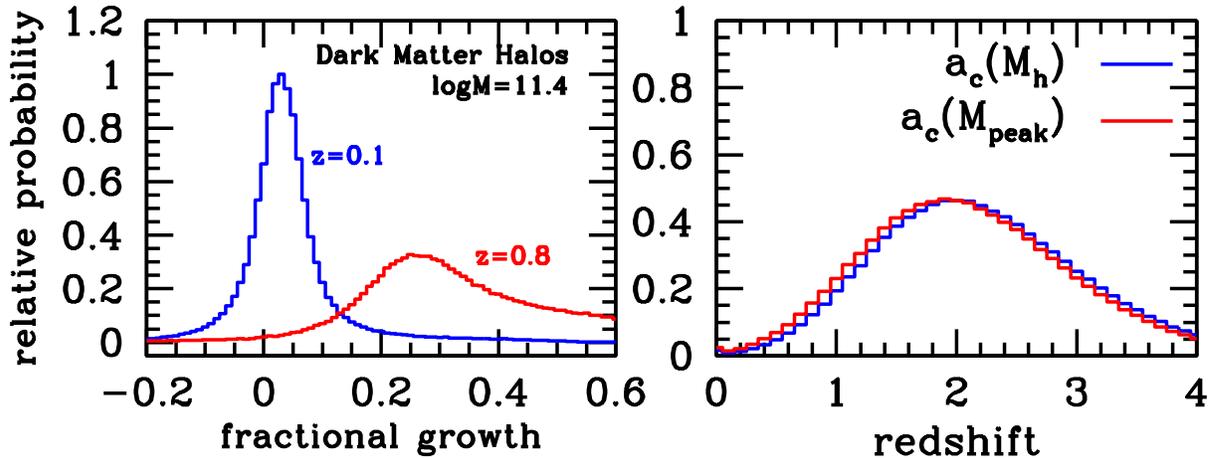,width=1\linewidth}
\vspace{-11cm}
\caption{ \label{mdot_histo} {\it Left Panel:} Fractional growth of
  halos from $z=0.1$ to 0 (blue histogram) and $z=0.8$ to 0 (red
  histogram) for halos of $\log\mhalo=11.4$. Over the short timeframe
  of $z=0.1\rightarrow 0$, a significant fraction of halos exhibit
  negative growth ($\sim 20\%$). This fraction is negligible for the
  $z=0.8\rightarrow 0$ timeframe. {\it Right Panel:} The distribution
  of halo formation epochs, as defined by the $a_c$ parameter of
  \citet{wechsler_etal:02}. The blue histogram shows the standard
  result when using halo mass as a function of time $(M_h)$. The red
  histogram shows the result when using $\mpeak(z)$ to determine halo
  growth. See text for details. }
\end{figure*}

\begin{figure}
\includegraphics[width=6.2in ]{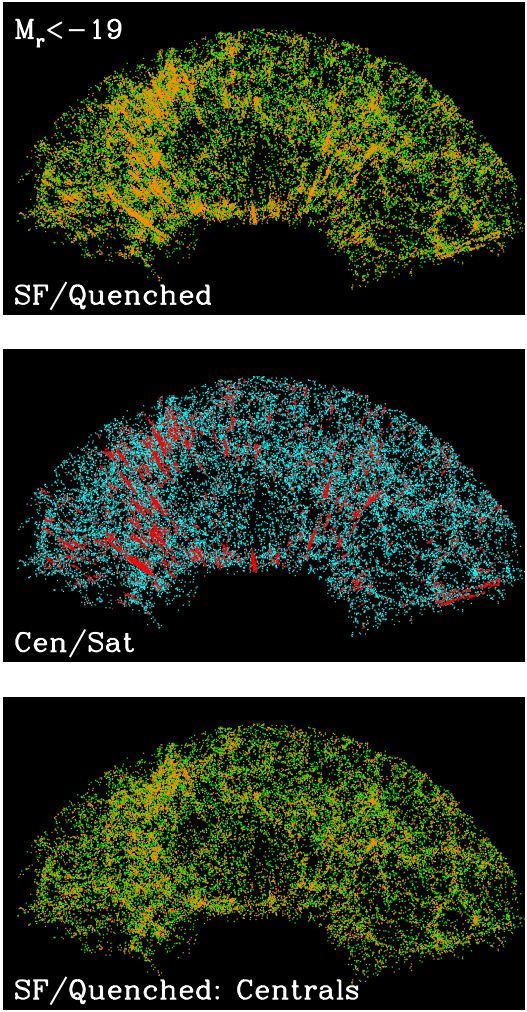}
\vspace{-1.5cm}
\caption{ \label{censat_slices} {\it Upper Panel:} A slice through a
  volume-limited sample of galaxies with $M_r<-19$. This sample
  extends to $z=0.064$. The color types of the points correspond to
  their star formation activity: the green points are on the
  star-forming main sequence while the orange points are quiescent
  galaxies on the red sequence. {\it Middle Panel:} The same set of
  galaxies, now categorized as central galaxies and satellite galaxies
  by the group finder. The group finder clearly identifies the
  fingers-of-god as redshifted galaxy groups, but some satellites
  exist in lower density environments. {\it Lower Panel:} The same as
  the upper panel, but now only central galaxies are being
  plotted. With the satellites removed, the finger-of-god effect is
  ameliorated, but there is a substantial fraction of quiescent
  galaxies, many of which reside in underdensities and voids. }
\end{figure}

\begin{figure}
\includegraphics[width=3.2in ]{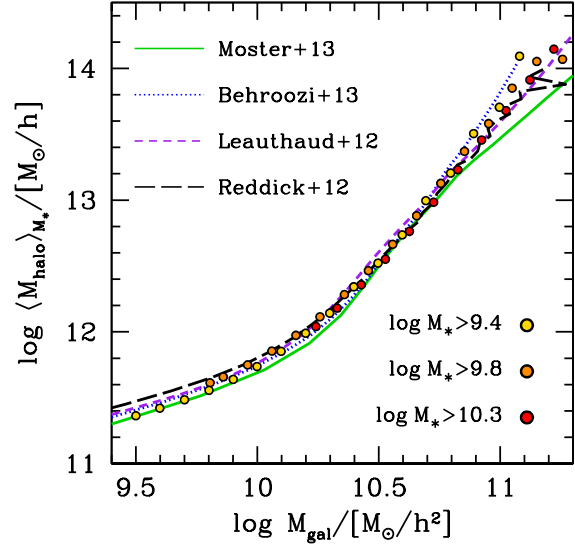}
\caption{ \label{mcentral} Average halo mass in bins of $\log\mgal$
  for central galaxies from the three volume-limited group
  catalogs. The four curves represent a sample of stellar to halo mass
  relation from abundance matching (\citealt{behroozi_etal:13,
    moster_etal:13}) and studies that combine abundance and clustering
  (\citealt{leauthaud_etal:12_shmr, reddick_etal:13}). }
\end{figure}

\section{Introduction}

The abundance matching model for connecting galaxies to halos has
proven to be an exceptional tool for understanding both galaxy bias
and galaxy evolution (see, e.g., \citealt{kravtsov_etal:04,
  conroy_etal:06, conroy_wechsler:09, moster_etal:10, moster_etal:13,
  behroozi_etal:13, reddick_etal:13}). In it's simplest form,
abundance matching places galaxies within halos based on their
relative ranking: the most massive galaxy goes in the most massive,
and on down the rank-ordered lists of galaxies and halos. The success
of abundance matching suggests a null hypothesis that galaxy
properties only care about the mass of their host halo. However,
correlations between galaxy and other halo properties at fixed halo mass
could manifest in spatial clustering; this is the well-known assembly
bias effect, in which halos of fixed mass cluster differently
depending on their formation history and internal structure
(\citealt{gao_etal:05, wechsler_etal:06, gao_white:07, wetzel_etal:07,
  li_etal:08, dalal_etal:08}). This idea has been tested in various contexts but with
conflicting results.

Using myriad galaxy clustering statistics, a number of studies found
no evidence for correlations of galaxy properties with environment at
fixed halo mass (\citealt{ abbas_sheth:06,
  skibba_etal:06,tinker_etal:08_voids}). In contrast, many results
using galaxy group catalogs to identify dark matter halos show that
galaxy properties depend on halo formation history as well as mass
(\citealt{yang_etal:06, wang_etal:08_assembly_bias, wang_etal:13,
  lacerna_etal:14}), provided there are no biases in the halo masses
induced by the group finding method
(\citealt{campbell_etal:15}). \cite{kauffmann_etal:13} find that star
formation rates of galaxies in separate halos are correlated, an
effect known as `galactic conformity'.  To model the conformity
results, \cite{hearin_watson:13} presented the `age-matching' model,
in which halos at fixed mass are rank-ordered by their formation time
and then abundance matched to color for the galaxies that occupy those
halos. Thus the oldest halos contain the reddest galaxies, while to
the youngest halos contain the bluest galaxies. A complication of
interpreting the age-matching model in the context of galaxy formation
is that `age' is quantity difficult to define objectively
(\citealt{li_etal:08}), and the redshifts at which halos accrue most
of their mass may not correlate (or may anticorrelate) with the
redshifts at which galaxies form or accrete most of their
stars. Additionally, \cite{geha_etal:12} find a limiting stellar mass
of $10^9$ $\hhmsol$ for field galaxies, below which no galaxies are
quenched. This represents a threshold below which halo formation
history can, by definition, play no role in whether a galaxy is
quenched because there are none, even though the amplitude of the
assembly bias effect only gets stronger as halo mass gets smaller.

In this series of papers, we test the assumption that halo growth and
galaxy growth are correlated. We construct this test in several
distinct regimes. In this paper, we focus on whether halo growth rate
correlates with whether a galaxy is quenched of its star formation and
resides on the red sequence. We use the spectral diagnostic $\dn$ to
separate galaxies into star-forming and quiescent samples. In a
companion paper, we test whether halo growth rate---as well as other
galaxy properties---correlates with galaxy star formation rate {\it
  within} the star-forming main sequence. Finally, this series will
also present new measurements of galactic conformity in the local
universe. To perform these tests, we use group catalogs created from
the NYU Value-Added Galaxy Catalog (\citealt{blanton_etal:05_vagc}),
which in turn were created from data from Data Release 7 of the Sloan
Digital Sky Survey (\citealt{sdss,dr7}). Results from the group
catalogs are compared to models created with high-resolution
cosmological N-body simulations. We will focus exclusively on
`central' galaxies---galaxies that reside at the center of distinct
halos, not orbiting within the virial radius of a larger halo. The
latter we classify as satellite galaxies. The formation histories of
central and satellite galaxies are quite different and are acted upon
by different physical mechanisms (see, e.g., \citealt{wetzel_etal:13}
and citations within). To isolate the effect of halo assembly bias on
the galaxy population, focusing on central galaxies makes this
comparison clearer.

Throughout, we define a galaxy group as any set of galaxies that
occupy a common dark matter halo, and we define a halo as having a
mean interior density 200 times the background matter density. A {\it
  host halo} is a halo that is distinct: its center does not reside
within the radius of a larger halo. We will use the terms {\it halo}
and {\it host halo} interchangeably in this work. A {\it subhalo} is
one whose center is located within the radius of a larger halo. For
all distance calculations and group catalogs we assume a flat, \lcdm\
cosmology of $(\om,\s8,\omb,n_s,h_0)=(0.27, 0.82,
0.045,0.95,0.7)$. Stellar masses are in units of $\hhmsol$. We will
sometimes refer to galaxies as `blue' and `red' to refer to their
intrinsic star formation; `red' means red-and-dead rather than red by
dust contamination.

\section{Data, Measurements, and Methods}

\subsection{NYU Value-Added Galaxy Catalog}

To construct our galaxy samples, we use the NYU Value-Added Galaxy
Catalog (VAGC; \citealt{blanton_etal:05_vagc}) based on the
spectroscopic sample in Data Release 7 (DR7) of the Sloan Digital Sky
Survey (SDSS; \citealt{dr7}).  We construct four volume-limited
samples that contain all galaxies brighter than $M_r-5\log h=-18$,
$-18.5$, $-19$ and $-20$, respectively. Within each
volume-limited sample, we determine the stellar mass at which the
sample is complete. The stellar masses are also taken from the VAGC
and are derived from the {\tt kcorrect} code of
\cite{blanton_roweis:07}, which assumes a \citet{chabrier:03} initial
mass function. In order of increasing luminosity, the stellar mass
sampels are complete at $\log \mgal=9.4$, 9.6, 9.8, and 10.3, where
stellar mass are once again in units of $\hhmsol$ (see Figure 2 in
\citealt{tinker_etal:11}).

For galaxy pairs that are too close to obtain spectra because of
the 55 arcsecond width of SDSS fibers (`fiber collisions'), we use the
internal correction to the fiber corrections within the VAGC, namely
that the collided object is given the redshift of the nearest galaxy
in terms of angular separation, provided that this redshift is in
agreement with the photometric redshift obtained by with the SDSS
photometry (\citealt{blanton_etal:05_vagc}).

Using galaxy color as a proxy for star formation activity can be
problematic, as dust reddening can cause a gas-rich disk galaxy to be
classified as a red sequence object
\citep{maller_etal:09,masters_etal:10_dust}.  To avoid this problem,
we use both $\dn$, which is a diagnostic of the light-weighted age of
the stellar population and thus is sensitive to the integrated star
formation history of the galaxy.  We obtain these quantities from the
JHU-MPA spectral reductions\footnote{\tt
  http://www.mpa-garching.mpg.de/SDSS/DR7/}
(\citealt{brinchmann_etal:04}).

\subsection{Measuring Large-scale Environment}
\label{s.galden}

For each galaxy, we estimate the large-scale environment by counting
the number of neighboring galaxies within a sphere of radius 10 \hmpc\
centered on each galaxy.  This quantity is a biased indicator of the
dark matter density field, but at 10 \hmpc\ this bias is a simple
linear factor and any stochasticity is minimal. We count the number of
galaxies above the corresponding magnitude threshold for the each
sample, and so the tracer of the density field has a different bias
for each sample. We do not correct for this between the samples, but
note that the relative bias between the different samples is at the
$\sim 5\%$ level (\citealt{swanson_etal:08_bias}). This galaxy density
measurement is affected by galaxy peculiar velocities, but this effect
is minimal at 10 \hmpc, as we demonstrate in Appendix A in
\cite{tinker_etal:11}. We also choose 10 \hmpc\ because represents a
clear distinction from a galaxy's small-scale environment as
encapsulated by its host halo. In tests we find that our results show
little dependence on the exact smoothing scale chosen. The mean number
of galaxies per 10 \hmpc\ sphere is 103, 72, 51, and 21 galaxies for
our volume-limited samples, going from faint to bright.

To correct for survey geometry and incompleteness, we use random catalogs. 
For each volume-limited sample, we produce a catalog of $10^7$ random
points distributed with the angular selection function of SDSS DR7
using the angular mask provided with the VAGC in combination with the
software package \texttt{mangle} (\citealt{swanson_etal:08}). Each
random point is also assigned a random redshift such that the comoving
space density of randoms is constant with redshift. For each galaxy,
we correct for incompleteness by multiplying the observed number of
galaxies by the ratio of the number of random points divided by the
expected number of randoms if the completeness were unity. The large
number of random points ensures that shot noise within each 10 \hmpc\
sphere is at the sub-percent level.

\begin{figure*}
\psfig{file=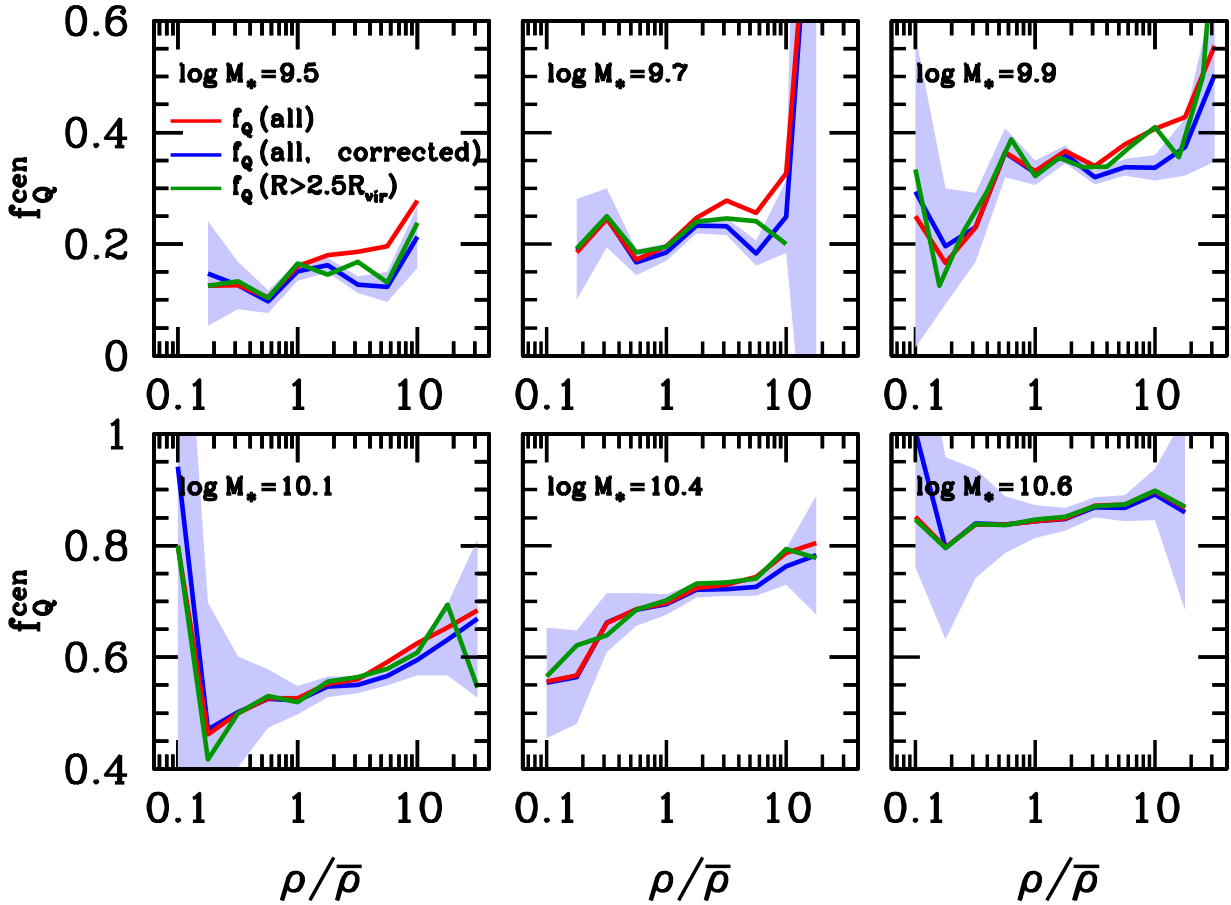,width=1\linewidth}
\vspace{-5cm}
\caption{ \label{fq_data} Measurements of the quenched fraction of
  central galaxies, $\fqcen$, as a function of large-scale
  density. Each panel shows three methods for measuring $\fqcen$: (1)
  The raw measurement that uses all centrals in the group catalog, (2)
  the measurement using all centrals in the group catalog, but
  statistically corrected for misidentification of centrals and
  satellites in the group-finding process, and (3) raw measurement
  that excludes centrals that are within $2.5\rvir$ of a larger
  group. The shaded region around the corrected measurement is the
  error in the mean, and is representative of the error on the other
  two measurements. At $\rho\la 1$, where the abundance of massive
  groups is low, all three measurements are essentially the same. At
  high densities and low stellar masses, the measurements separate at
  high densities. For high stellar mass bins, the frequency of groups
  more massive than the host halos being probes is small, and all
  three measurements are consistent at all $\rho$.
}
\end{figure*}

\begin{figure*}
\psfig{file=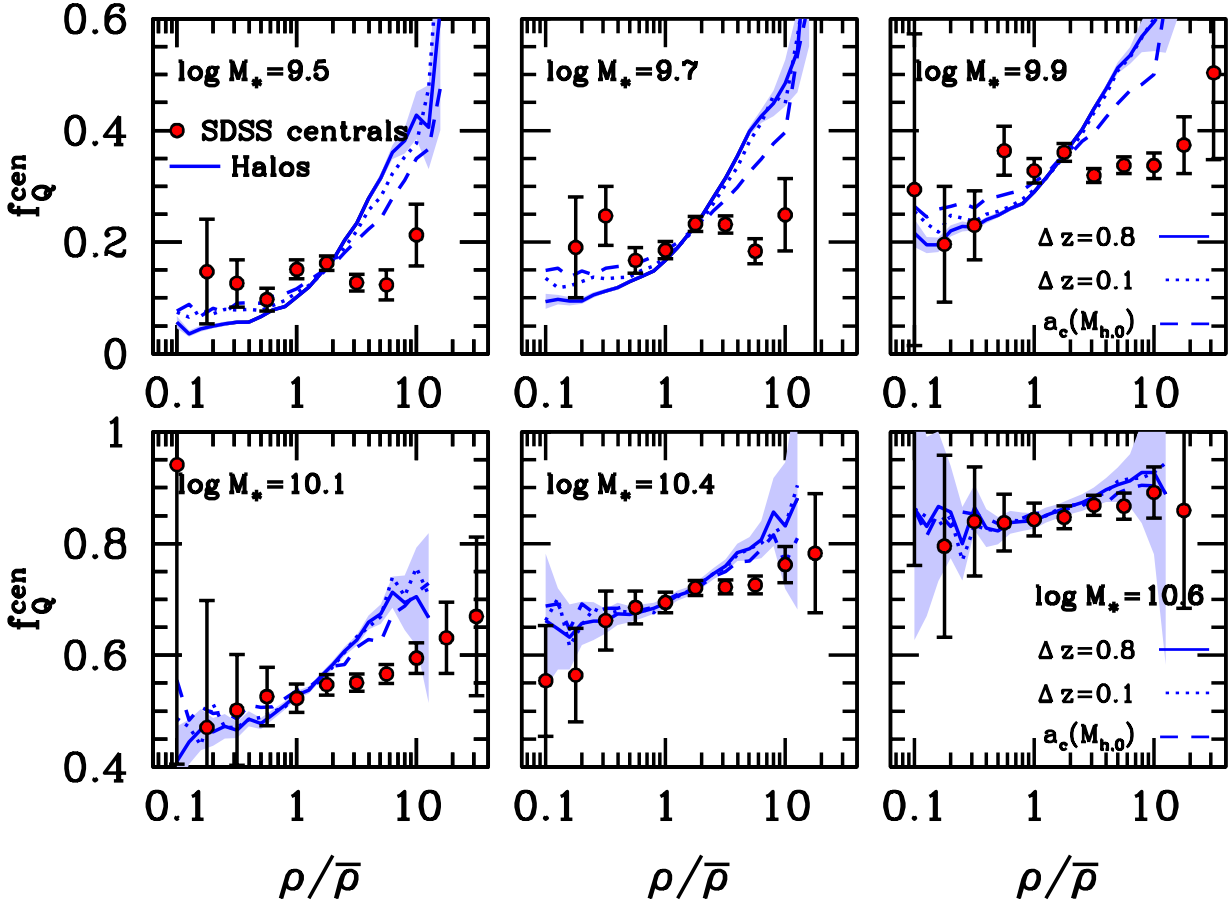,width=1\linewidth}
\vspace{-5cm}
\caption{ \label{mdot_delta} The quenched fraction of central
  galaxies, $\fqcen$, as a function of large-scale galaxy density, for
  six bins in $\mgal$. Galaxy density measurements are described in \S
  \ref{s.galden}. Error bars are the error in the mean in each $\rho$
  bin. In each panel, we compare these measurements to expectations
  from the age-matching model, which stipulates that redder galaxies
  (or, in these data, galaxies with the largest values of $\dn$, which
  implies that they have the oldest stellar populations) live in older
  halos. We use the group catalog to estimate the halo masses for each
  bin in stellar mass. In each bin in halo mass, the halos are
  rank-ordered by three different definitions of age: (1) their
  fractional growth since $z=0.8$, (2) their fractional growth since
  $z=0.1$, and (3) their formation epoch as defined by $a_c(M_h)$. We
  set the break point between `old' and `young' halos to match the
  value of $\fqcen$ in each bin. As expected from halo assembly bias,
  the old fraction of halos depends strongly on large-scale
  environment. The assembly bias gets less strong monotonically with
  increasing $\mgal$. The data, in contrast, show the opposite
  trend. At low stellar masses, there is little to no dependence of
  $\fq$ on environment. As $\mgal$ increases, $\fq$ shows a positive
  trend with $\rho$.  This figure is an updated version of one
  presented in \citealt{tinker_etal:11}, with new simulation
  predictions and correcting for an error in the density calculations
  around the galaxies. }
\end{figure*}

\begin{figure*}
\psfig{file=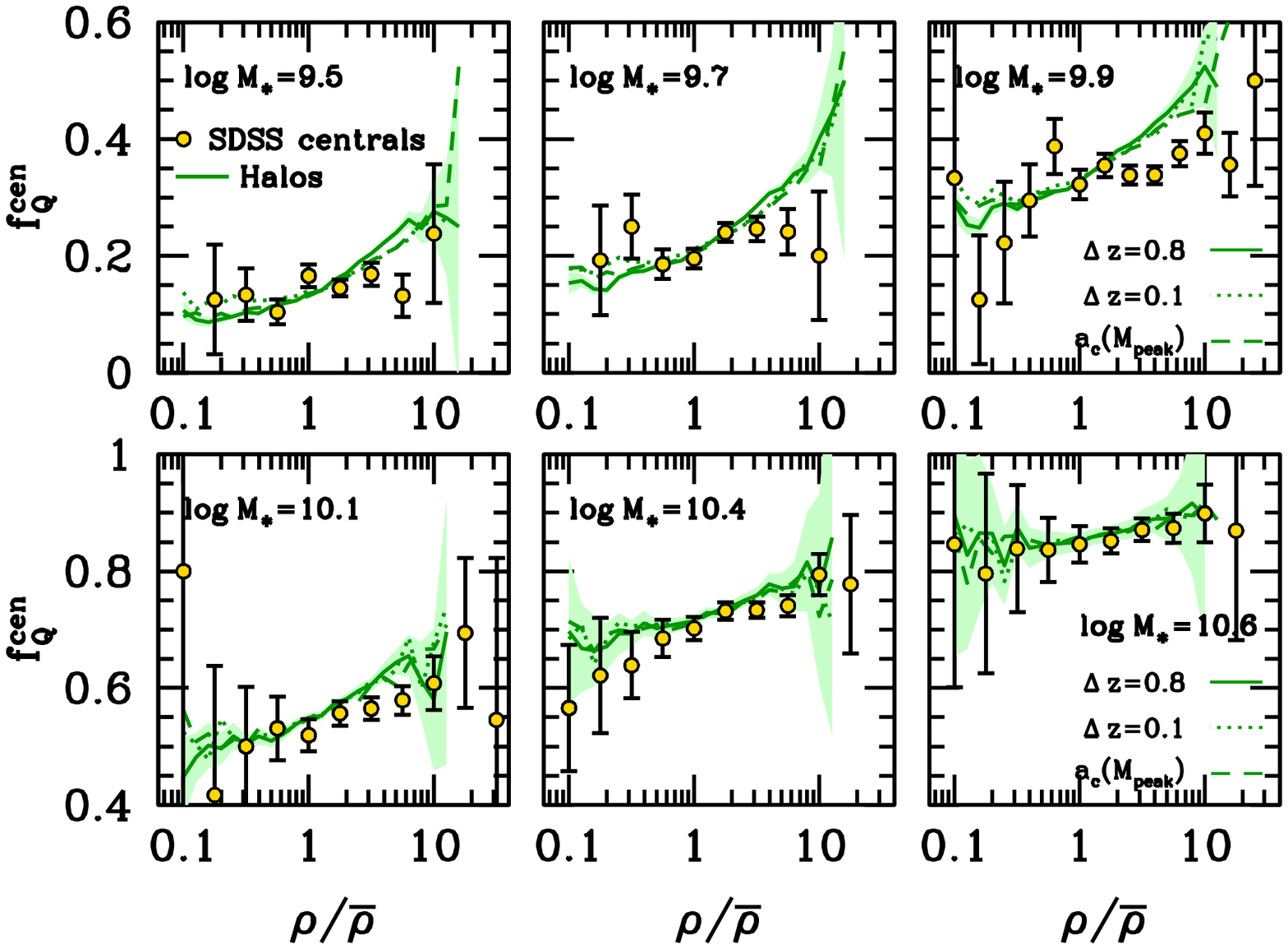,width=1\linewidth}
\vspace{-5cm}
\caption{ \label{mdot_delta_nosplash} Analogous to Figure
  \ref{mdot_delta}, but removing galaxies that are near larger halos
  to eliminate the possible effects of splashback encounters.  For the
  data, we remove from the sample any central galaxy that is within
  $2.5\rvir$ of a larger halo. The data look nearly the same as in
  Figure \ref{mdot_delta}, but recall that those data have been
  statistically corrected for bias in $\fq$, while these data have not
  been corrected in any way. The three curves show the predictions of
  the old fraction of dark matter halos. We use the same three
  definitions of halo age as before. However, for $\Delta z=0.8$ and
  $\Delta z=0.1$, we remove all halos that are within $2.5\rvir$ of a
  larger halo, just as with the data. For formation epoch, $a_c$, we
  use $\mpeak$ rather than $\mhalo$, but use all halos in the
  sample. This removes the impact of tidal events and splashbacks
  without removing any halos.}
\end{figure*}

\subsection{Group Finding Algorithm}

We use the halo-based group-finding algorithm presented in
\cite{tinker_etal:11}, which is, in turn, based on the algorithm of
\cite{yang_etal:05}. In brief, the group finder uses the abundance
matching ansatz to assign halo masses to groups, iterated until
convergence. The resulting group catalog is a robust decomposition of
the entire galaxy population into central galaxies and satellite
galaxies. This group finder has been thoroughly vetted in
\cite{tinker_etal:11} as well as \cite{campbell_etal:15}, which
specifically investigated color-dependent statistics derived from the
group finders. \cite{campbell_etal:15} concluded that our group finder
can robustly identify red and blue centrals and satellites as a
function of their stellar mass, but the assignment of halo masses is
highly problematic. Thus when using the group catalog, we will only
divide the results based on $\mgal$ and not $\mhalo$.

\subsection{Numerical Simulations and Defining Halo Growth}
\label{s.simulations}

We compare the results from the group catalog to expectations from
dark matter halos. We use the `Chinchilla' simulation (Becker
et.~al.~in prep.), run using a varient of the Gadget-2 cosmological
N-body code (\citealt{springel:05}) known as L-Gadget2. The box size
is 400 \hmpc\ per side, evolving a density field resolved with
$2048^3$ particles, yielding a mass resolution of $5.91\times 10^8$
\hmsol. The cosmology of the simulation is flat \lcdm\ consistent with
recent CMB results, with $\om=0.286$, $\sigma_8=0.82$, $h=0.7$, and
$n_s=0.96$. This is slightly higher matter density than assumed for
the group catalogs, but the change makes negligible difference in any
comparison.

Halos are found in the simulation using the Rockstar code of
\cite{rockstar}. Halos masses are defined as spherical overdensity
masses according to their virial overdensity. We use Consistent Trees
(\citealt{consistent_trees}) to track the merger and growth history of
each halo in the simulation, and we use these histories to determine
the growth rate of each halo.

Figure \ref{mdot_histo} shows the fractional growth of dark matter
halos at low mass over different redshift baselines. Halos of this
mass are likely to contain galaxies of $\log\mgal=9.4$. Detecting
assembly bias in survey data has a particular challenge: assembly bias
usually increases with decreasing halo mass, but smaller mass halos
contain dimmer galaxies are can only be seen in small volumes. The
choice of $\log\mhalo=11.4$ represents a compromise between these two
effects: halo assembly bias seen in simulations is significant, but
SDSS still probes these galaxies at cosmological volumes. 

Over the redshift baseline $z=0.8\rightarrow 0$, halos of this mass
scale grow on average $\sim 30\%$, but with a wide
distribution. Notably, the fraction of halos with negative growth over
this timeframe is negligibly small. Over the timeframe of
$z=0.1\rightarrow 0$, the variance in halo growth rates is much
smaller, but roughly a quarter of halos lose some mass. Some of this
is noise\footnote{We estimate the noise in assigning halo masses in
  the Rockstar code by calculating the snapshot-to-snapshot variance
  around the mean trend in halo growth for each halo for the five
  snapshots that cover the redshift range $z=0.1\rightarrow 0$. The
  variance depends on the order of the polynomial used to fit for the
  mean trend in $\mhalo(z)$, but for a second-order polynomial the
  variance is 1.3\%. We conclude that this is an upper limit on the
  noise in estimating halo mass. The variance in the fractional growth
  is 4\%. Assuming Gaussian statistics, removing the contribution from
  noise would reduce this only to 3.6\%.}, but these halos exhibit the
strong clustering indicative of assembly bias (as we will see in \S
3), indicating that noise is a minority contributor. Hereafter, we will
refer to these redshift baselines as `$\Delta z$'. For $\Delta z=0.8$,
this definition is over an intermediate timespan ($\sim 7$ Gyr) over
which most of these central galaxies arrive on the red sequence
(\citealt{tinker_etal:13_shmr}). For $\Delta z=0.1$, this definition
is sensitive to short-term growth of the dark matter halos. To
implement the age-matching model using these definitions, halos are
rank ordered from lowest to highest fractional growth. The lowest are
considered the oldest, while the halos that have grown the most over
that timespan are considered the youngest.

In the right-hand panel, we show the `age' of the same halos as
defined by the $a_c$ parameter from \cite{wechsler_etal:02}, which
identifies the epoch where halo growth changes from rapid accretion to
slower growth. This is qualitatively similar to using half-mass epoch,
although the median of $z(M_{1/2})$ is lower than that of the median
redshift of $a_c$. When we quantify assembly bias of halos, we find
little difference between using $z(M_{1/2})$ and $a_c$, thus we will
focus of $a_c$. The two histograms in this panel show $a_c$ for two
different definitions of halo mass: (1) the current halo mass at any
time, $M_h(z)$, and (2) the peak halo mass up to that time $M_{\rm
  peak}(z)$. As shown in the left-hand panel, over short time
intervals there can be significant dark matter mass loss. In fact,
small halos can be accreted onto a larger halo but have too much
kinetic energy to remain within the larger halo, eventually exiting
the larger halo after one pericentric passage. These are called
`splashback' halos, and this process can lead to significant stripping
of the dark matter halo. $\mpeak(z)$ is a monotonically increasing
function, thus for halos that experience tidal stripping, $M_h(z)$
will be smaller than $\mpeak(z)$. This has a small but visible impact
on the distribution of formation epochs, pushing them to slightly
smaller redshift. When rank-ordering halos by their age (or fractional
growth), halos that have encountered significant tidal encounters get
pushed to the top of the list. Thus, in the standard age-matching
model, these halos house the oldest galaxies. Using $\mpeak$
effectively removes the impact tidal events or splashback galaxies on
the ordering of the list. The overall effect on the distribution of
halo ages is small, but as we will see in the following section, this
choice has a major impact on the predicted assembly bias.

To compare simulation results to galaxy results binned as a function
of environment, we measure the density around each halo in the
simulation in the same manner as for the galaxies. Using the
halo occupation distribution (HOD) fitting results of
\cite{zehavi_etal:11} from the SDSS Main galaxy sample, we populate
the simulation with galaxies that match the density and clustering of
each of our volume-limited samples. Using the distant-observer
approximation and the $z$-axis of the box as the line-of-sight, the
top-hat redshift-space galaxy densities are measured around each halo. 

In an appendix we show the results of two additional proxies for halo
age: the redshift at which a halo reaches half its present-day mass,
$\zhalf$, and halo concentration, $\cvir$, which has been shown to
correlate tightly with formation history (\citealt{wechsler_etal:02})
and is one of the primary quantities through which halo assembly bias
manifests. We do not include these in the main text as they are
quantitatively similar to the definitions already in hand. In this
appendix, we also show the `break points' delineating old and young
halos.

\begin{figure}
\psfig{file=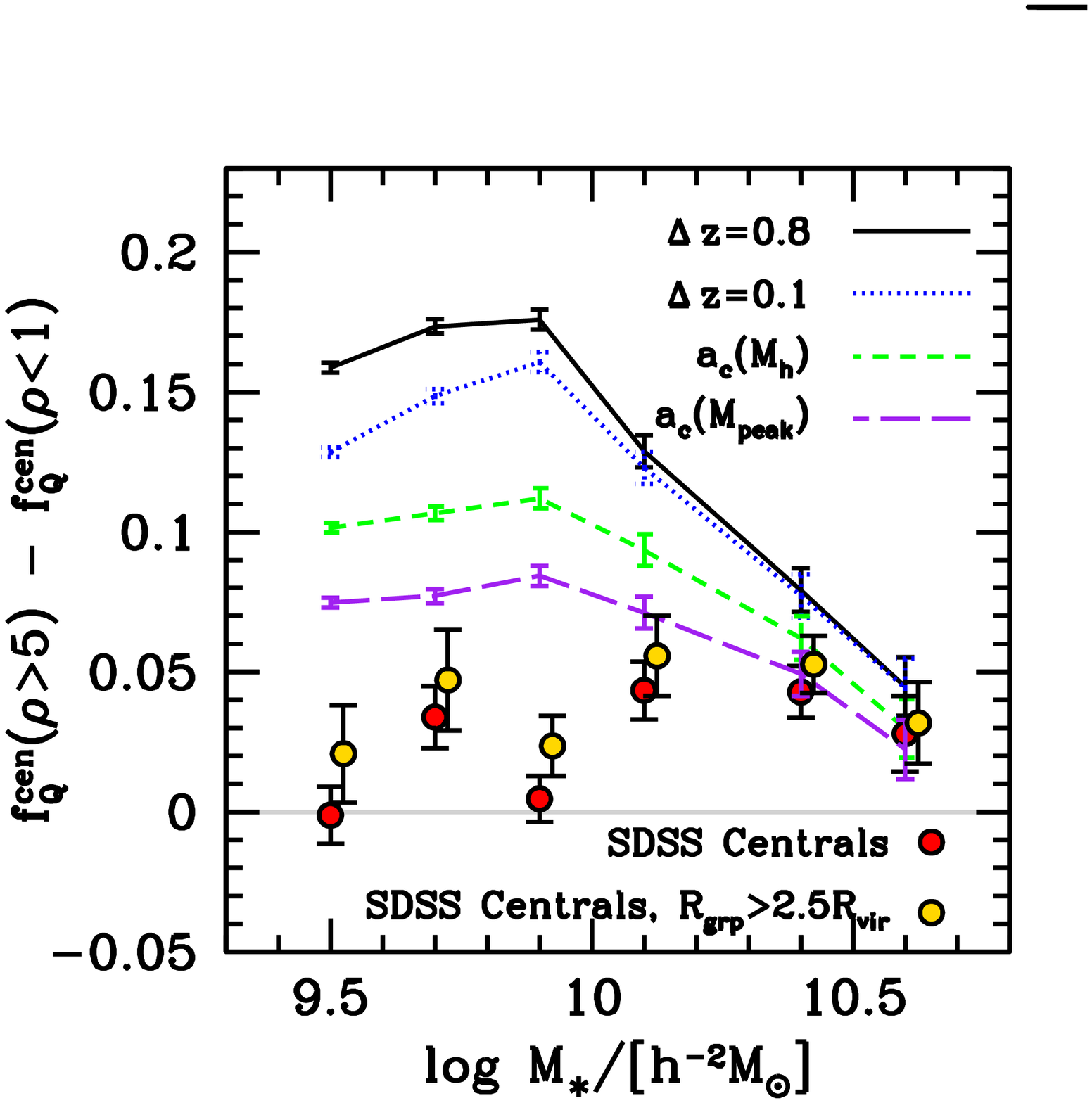,width=1.05\linewidth}
\caption{ \label{delta_fq} Assembly bias in SDSS central galaxies and
  in dark matter halos. The $y$-axis is the difference between the
  quenched fraction of central galaxies at high densities ($\rho>5$)
  and low densities ($\rho<1$). The red and yellow circles represent
  SDSS centrals for the full sample (red) and when galaxies near
  larger groups have been removed (yellow). Recall that the red points
  have been corrected for biases in the group catalog. Error bars are
  error in the mean. Curves show predictions from dark matter halos
  for four different definitions of halo age. We do not show any
  predictions from age-matching models where halos near groups are
  removed because they are consistent with the results from the
  $a_c(\mpeak)$ model. }
\end{figure}

\section{Results}

\subsection{Central-satellite decomposition of the SDSS}

Figure \ref{censat_slices} shows an example of the group finder
applied to one of our volume-limited samples. The top panel shows
galaxies in the NGC footprint (only plotting galaxies with
$\delta<22^\circ$ to avoid crowding), where the color of the point
indicates whether the galaxy is star-forming or quiescent. The middle
panel shows the same galaxies, but now color indicates whether the
galaxy is a central or a satellite. The group catalog clearly
identifies the `fingers-of-god' created by the large virial motions of
satellite galaxies. Satellites are mostly found in high-density
regions along filaments in the cosmic web, although some satellites
are still found occasionally out in the field. The bottom panel shows
the star-forming and quiescent breakdown of the sample, but now for
central galaxies only. Although satellites are more likely to be on
the red sequence than central galaxies, this panel elucidates two
aspects of central galaxies: (1) the overall fraction of central
galaxies on the red sequence is significant at these
masses, and (2) that
quenched central galaxies exist at all densities, even the deepest
void in the galaxy distribution.

Figure \ref{mcentral} shows the relationship between $\mhalo$ and
$\mgal$ for central galaxies for three volume-limited group
catalogs. The results between the catalogs are in excellent
agreement with one another, as well as with results from the
literature constraining this relationship from different methods. We
use this relationship to make subsequent comparisons between our halo
catalogs and the SDSS measurements. Although the mean $\mhalo$ in bins
of $\mgal$ is not equivalent to the inverse relationship, the
differences between Figure \ref{mcentral} and its inverse only appear
at $\mgal \ga 10^{10.6}$, above the limit for our comparisons.

\subsection{Quenched fraction of central galaxies and large-scale density}

The correlation between density and galaxy properties is well-known
(see \citealt{blanton_moustakas:09} and for a thorough
review). Progressing from low to high densities, the fraction of
galaxies that are red-and-dead, $\fq$, monotonically grows (see, e.g.,
\citealt{oemler:74, davis_geller:76, dressler:80} for canonical works
and \citealt{hogg_etal:04, kauffmann_etal:04, blanton_etal:05a,
  baldry_etal:06, park_etal:07, bamford_etal:09} for more recent
measurements). However, this observation combines central galaxies
that live in low-mass dark matter halos with satellite galaxies that
orbit within high-mass dark matter halos. The observed trend is driven
by the density dependence of the halo mass function: more massive
halos live in more dense environments, and in turn have a higher
fraction of quenched galaxies. The question we pose here is, when
restricting the sample to central galaxies of fixed stellar mass,
which is a reasonable proxy for fixed halo mass, what is the
correlation between $f_Q$ and environment?

We define red-and-dead galaxies as those with $\dn>1.6$. We find
that this value of 1.6 faithfully follows the minimum of the
distribution of $\dn$ values between the SFMS and the red sequence,
irrespective of galaxy stellar mass. We also note that the results are
nearly indistinguishable when using sSFR as our indicator of
quiescence. Figure \ref{fq_data} shows the quenched fraction of
central galaxies, $\fqcen$, as a function of large-scale galaxy
density, measured in three different ways: 

\begin{itemize}  
\item (1) We present the raw
measurements of $\fqcen(\rho)$ in which all central galaxies in the
group catalog are used. 
\item (2) We apply a statistical correction to
$\fqcen(\rho)$ to remove biases imparted by the group-finding process. 
\item(3) We measure $\fqcen(\rho)$ after removing all
central galaxies that are within $2.5\rvir$ of a larger group.
\end{itemize}

For (2), $\fqcen$ is corrected for impurities in the group catalog as
in Appendix C in \cite{tinker_etal:11}. In the group-finding process,
centrals and satellites are sometimes misclassified, leading to $\sim
10\%$ of central galaxies in the catalog being true satellites. This
effect increases $\fqcen$ because satellite galaxies always have
higher quenched fractions at fixed $\mgal$, and the misclassification
occurs more frequently in higher density regions that contain more
satellites. The statistical correction described in
\cite{tinker_etal:11} is applied directly to measurements of $\fq$,
and robustly accounts for biases in the group-finding process. For
$\mgal<10^{10}$ $\hhmsol$, the corrected and uncorrected measurements
of $\fqcen$ at low densities are the same. This is expected because
the abundance of satellite galaxies is negligible at $\rho<1$. At
higher $\rho$, the correction factor lowers $\fq$ by roughly 0.05 to
0.10, also expected from the amount of misclassification in the group
catalog (Appendix C in \citealt{tinker_etal:11}). At higher stellar
masses, the corrected and uncorrected results are consistent at all
$\rho$ due to the fact that there is a smaller difference in the
quenched fractions of central and satellite galaxies than for lower
$\mgal$.

For (3), as discussed in \S \ref{s.simulations}, halo growth can be
negative. How this impacts the growth of galaxies is not fully
understood, but splashback galaxies are subject to environmental
processes, such as ram pressure and strong tidal stripping, that more
isolated galaxies are not subjected to. Thus they are not clean tests
of the correlation of halo growth history to galaxy formation, and it
makes sense to treat them as a separate class of galaxies.
\cite{wetzel_etal:14} showed that splashback galaxies essentially
behave the same as satellite galaxies, meaning that after the initial
accretion event their evolution is unchanged for a long delay time,
and then they rapidly quench their star formation. The key quantity is
the time of the initial accretion event, regardless of the previous
evolutionary history of the halo; i.e., regardless of whether it was
an early-forming halo or late-forming halo before the accretion event.
This then begs the question: If splashback galaxies and halos are
removed from consideration, what are the observations and theoretical
predictions?

To implement (3), we remove all central galaxies with projected
separation $R<2.5\rvir$ of a larger group and $\Delta v<1000$ km/s
with respect to the central galaxy of the larger group.  These choices
are motivated by the results of \cite{wetzel_etal:14} and references
therein. The measurements of $\fqcen$ after this process split the
difference between the raw measurements and the corrected
measurements.  The statistical correction is not applied here because
it is only applicable on the full sample of galaxies. But in
comparison to the raw data using all centrals, removing galaxies near
groups lowers $\fq$ in high densities and low stellar masses. For
$\mgal\ga 10^{9.8}$ there is little difference between the raw
measurements and those with no galaxies near groups.

Regardless of how $\fqcen$ is measured, the results indicate that, at
$\mgal<10^{10}$ $\hhmsol$, there is little correlation between quenched
fraction and large-scale environment. At $\mgal>10^{10}$ $\hhmsol$,
there is a shallow but significant positive slope of $\fqcen$. We will
compare these results to predictions of the age-matching model in the
following sections.

\subsection{Does halo growth correlate with a galaxy being on the red
  sequence?}

Figure \ref{mdot_delta} shows the measurements of $\fqcen$, using all
central galaxies and corrected for group-finding biases. The curves
show the prediction of the age-matching model, which puts the oldest
galaxies into the oldest halos (once fixing halo mass). We choose
$\mhalo$ by the halo mass assigned to galaxies in each bin, but we
note that the halo predictions vary weakly with halo mass (as can be
seen in the figure). Thus, minor biases in $\mhalo$ and scatter
between halo mass and stellar mass are not likely to change the
predictions. Halos are rank-ordered by three of the metrics for halo
age presented in \S \ref{s.simulations}: halo growth over the redshift
ranges $\Delta z=0.8$ and $\Delta z=0.1$, and the formation epoch
$a_c(M_h)$. We set the break point between `old' and `young' halos
such that the old fraction of halos matches the observed $\fqcen$ in
the data.

The age-matching curves in Figure \ref{mdot_delta} indicate how the
fraction of old halos depends on $\rho$. Under the age-matching
hypothesis-- regardless of age definition---the old fraction has a
strong dependence on $\rho$, with the majority of old halos living in
dense environments. This result is consistent with previous results of
halo assembly bias, but at odds with the observations at most stellar
mass bins, most notably at lower stellar masses. Even at
$\mgal>10^{10}$ $\hhmsol$, where there is a measurable trend of
$\fqcen$ with $\rho$, the standard age-matching model predicts a
correlation stronger than that seen in the data. At lower masses, the
data and theory are at loggerheads: the strength of the assembly bias is at
its largest, but the data show the weakest correlation between
$\fqcen$ and $\rho$, if at all.

\subsection{Removing the impact of splashback halos and galaxies}

Figure \ref{mdot_delta_nosplash} shows analogous measurements and
models as Figure \ref{mdot_delta}, only here, splashback effects have
been removed. For the SDSS central galaxies, any galaxy that is within
a projected separation of $2.5\rvir$ of a larger halo, along with
$\Delta v<1000$ km/s, is removed from the sample. From
\cite{wetzel_etal:14}, this will remove most all splashback galaxies.

The curves in each panel represent the predictions of the age-matching
model using the same three definitions of halo age as before. For
$\Delta z=0.8$ and $\Delta z=0.1$, the halo samples have been altered
in the same fashion as the data: all halos within $2.5\rvir$ of a
larger halo have been removed from the sample. In comparison to the
age-matching predictions of Figure \ref{mdot_delta}, the assembly bias
signal is substantially reduced: there is a definite trend of higher
old fraction in higher densities, but not nearly as steep as the trend
for all halos. The final age-matching prediction, using $a_c$, but now
$a_c$ is defined using $\mpeak$ rather than $\mhalo$. For this model,
{\it no halos are removed} from the catalog. We include this model
here, rather than Figure \ref{mdot_delta}, to show that the
$a_c(\mpeak)$ model induces the same level of assembly bias as models
that remove all potential splashback effects.

At lower masses, the age-matching model predicts a correlation of
$\fqcen$ with $\rho$ not seen in the data, although the differences
between theory and data are smaller than that seen with the standard
age-matching implementation. At higher masses, $\mgal>10^{10}$
$\hhmsol$, there is reasonable agreement between the age-matching
models and the positive trend of increasing $\fqcen$ with $\rho$. 

\subsection{Assembly Bias in Halos and Galaxies}

Figure \ref{delta_fq} summarizes our results on assembly bias in
galaxies and halos. The $y$-axis shows the difference between the
quenched fractions at high and low densities,
$\fqcen(\rho>5)-\fqcen(\rho<1)$, as a function of stellar mass. At low
masses, the results are generally consistent with little to no
assembly bias. At high masses, there is a statistically robust
assembly bias signal, with red fractions being around 0.05 higher at
high densities. The overall quenched fraction at these masses
approaches unity, thus another way to phrase the result is that the
{\it blue fraction} of central galaxies in high densities is 10-20\%
lower than in lower densities.

The curves show the predictions of the age-matching models for four
different age definitions. The first three show the models from Figure
\ref{mdot_delta}, in which all halos are used at each mass bin. The
fourth model uses $a_c(\mpeak)$ as the halo age definition, although
we note that all the theoretical models from Figure
\ref{mdot_delta_nosplash} are consistent with one another. At low
stellar mass, Figure \ref{delta_fq} conveys two important points: 1)
that the amplitude of the assembly bias signal depends strongly on how
one defines halo age, and 2) that none of these models are in
particularly good agreement with the data. At high stellar masses, the
comparison of halos and galaxies is quite different. There is still a
dependence of the assembly bias signal on age definition, but the
prediction of the $a_c(\mpeak)$ model is in reasonable agreement with
the data. We note that this implies that all models that remove
possible splashback halos will also be in agreement.

\section{Discussion}

The main results of this paper are:

\begin{itemize}
\item The predictions of the age-matching model depend on how age is
  defined. More specifically, once tidal and splashback effects are
  removed from consideration by use of $\mpeak(z)$ rather than
  $\mhalo(z)$, the amount of assembly bias is reduced. This has the
  largest effect on low-mass halos $\mhalo\lesssim 10^{12}$ \hmsol. 
\item At low galaxy mass, $\mgal\lesssim 10^{10}$ $\hhmsol$, the results
  are consistent with litte-to-no assembly bias, implying no
  relationship between halo age and galaxy quenching for central
  galaxies.
\item At higher galaxy masses, the results are consistent with predictions
  from the age-matching model after removing the effects of splashback
  halos by using $\mpeak(z)$ to characterize formation history.
\end{itemize}

The first point is important for properly framing our expectations
from assembly bias and the age-matching model. The extreme assembly
bias predictions at low masses are driven by tidal and splashback
effects lowering the present-day mass of the halo relative to its peak
value at some earlier time. \cite{wetzel_etal:14} shows that the
galaxies within these halos are not immediately affected by these
encounters: after accretion onto a larger halo, the galaxy evolves as
though it were still in the field for 3-5 Gyr. Most halos will be
reaccreted by the larger halo during that time, and all will
eventually be reaccreted onto the larger halo. There is a measurable
increase in the quenched fraction within a couple virial radii of a
larger halo, but splashback galaxies are not a significant contributor
to the whole population of central galaxies on the red sequence. Thus
standard age-matching predictions generally overestimate the impact of
halo formation history on galaxy quenching. One aspect of the
age-matching model that has received little attention (and this work
is no different) is the possibility of scatter in any halo age-galaxy
age correlation. A one-to-one correspondence between these two
properties is unlikely, and scatter is a key component of the standard
abundance matching model. Scatter between halo age and galaxy age
would reduce the amplitude the assembly bias in the galaxy
population. It is possible that a physically reasonable amount of
scatter could reconcile the standard age-matching model with
observations at high $\mgal$. More work is required to define
`physically reasonable scatter', but at low masses the amount of
scatter required to bring age-matching into agreement with the data
would be so large as to eliminate any effective correlation.

The measurements of $\fqcen(\rho)$ cannot be reconciled with the
predictions of the age-matching model at low $\mgal$. This result caps
a number of other results that are mutually exclusive with a model
that maps halo age onto galaxy age at these mass
scales. \cite{tinker_etal:08_voids} demonstrated that the sizes of
voids in red and blue galaxies is consistent with galaxy color being
independent of large-scale environment, and inconsistent with the
level of assembly bias seen in red galaxies in, for example, the
\cite{croton_etal:07} semi-analytic model. When comparing the results
of age-matching models to measurements of galaxy clustering and
galaxy-galaxy lensing, there are conflicting results in the
literature. \cite{hearin_etal:14} show reasonable agreement between
the age-matching model and measurements of color-dependent clustering
and lensing. In contrast, \cite{mandelbaum_etal:16} and
\cite{zu_mandelbaum:16} find that the predictions of the standard
age-matching model are inconsistent with galaxy-galaxy lensing
measurements split by color in bins of galaxy stellar mass. One
difference between these two analyses is that Hearin et.~al.~compare
models to data in thresholds of stellar mass, while the other papers
compare modela and data in bins of stellar mass.
Additionally, \cite{zehavi_etal:11} use a standard halo occupation
formalism to fit the color-dependent clustering of SDSS galaxies in
multiple, narrow bins of color at fixed galaxy luminosity. In the
standard HOD approach, galaxies occupy halos based only on the halo
mass. Thus, if color depended significantly on halo age at fixed mass,
the standard HOD approach would not be able to fit the clustering
data. \cite{zentner_etal:14}, using mock galaxy samples that contain
assembly bias, do obtain a good fit to mock clustering using the
standard HOD approach, but the clustering was measured in threshold
samples, not magnitude bins as done in \cite{zehavi_etal:11}.

The third point above implies that there is some change in how
quenching correlates with halo formation history between low stellar
masses and high stellar masses. Either that change manifests from a
change in the physical mechanism that quenches galaxies, or that the
mechanism is the same but the correlation between that mechanism and
halo formation history---i.e., the scatter discussed above---increases
significantly as halo mass decreases. \cite{dalal_etal:08} show that
the assembly bias in low and high mass halos, split around $\mhalo\sim
10^{12}$ \msol, is caused by different physical mechanisms. As we have
noted above, assembly bias in low mass halos is driven by tidal
encounters and other interactions with the large-scale
environment. Assembly bias in high-mass halos, on the other hand, is
imprinted in the primordial density field; early- and late-forming
halos can be identified by the nature of their initial
perturbations. The process that quenches low-mass field galaxies must
be nearly independent of environment and thus uncorrelated with halo
formation history. This does not necessarily imply that all the
properties of low-mass field galaxies are uncorrelated with the
details of halo growth; star formation rates, galaxy sizes, and
morphologies may correlate with short term or long term halo growth
rates. The results here only indicate that the decision to migrate
from the star-forming sequence to the red sequence is not up to the
halo, after accounting for halo mass.

In contrast, high-mass galaxies tell a different story.
\cite{tinker:16} shows that, if quenching is induced by a threshold in
either galaxy mass or halo mass, the epoch of quenching will depend on
halo formation history, with early-forming halos quenching
earlier. \cite{tinker_etal:12_cosmos} found that the clustering of
x-ray groups depended on the state of the central galaxy; groups with
star-forming centrals had higher clustering at $z\sim 1$. Halo
assembly bias has also been, for the first time, robustly detected in
observations of cluster sized halos (\cite{miyatake_etal:16,
  more_etal:16}). The picture these results paint is consistent with
the results here; older massive halos are more likely to contain
quenched galaxies than younger halos, but the overall size of the
effect is relatively small compared to the mean quenched fraction of
high-mass galaxies.

We will tackle galactic conformity in a future paper in this series,
but the results presented here are inconsistent with a model in which
assembly bias creates strong large-scale galactic conformity for
low-mass galaxies (i.e., conformity outside the virial radii of the
halos in which the galaxies lie). There are, however, different
definitions of galactic conformity that can be lead to different
quantitative results. In this paper, we focus on the quenched fraction
of central galaxies, the same as the conformity definition used by
\cite{hearin_etal:15}. \cite{kauffmann_etal:13} measure conformity by
measuring median star formation rates around samples of large isolated
galaxies, where the isolated galaxies are divided into many bins based
on their specific SFR. \cite{kauffmann_etal:13} find a strong
suppression of sSFR of galaxies around the least star-forming isolated
galaxies. The stellar mass range at which \cite{kauffmann_etal:13}
find conformity is consistent with the stellar masses at which we find
a weak trend of $\fq$ with $\rho$. It's possible to change the mean
sSFR without altering $\fq$, thus further study is required to see if
these observations are compatible. Additionally, a separate effect
known as small-scale galactic conformity---the properties of satellite
galaxies conforming to that of the central galaxy, first detected by
\cite{weinmann_etal:06}---has been confirmed by other studies
(\citealt{knobel_etal:15, kawinwanichakij_etal:16, berti_etal:16}).

A robust theory of galaxy formation must be consistent with all of
these results listed above: a formation path that yields clear
conformity within a dark matter halo, conformity of star formation
rates outside of the halo, but limited to no correlation of the
quiescent fraction on large-scale environment. A convincing
explanation for all these observations will likely combine the
influence of dark matter structure formation with complicated
astrophysics and phenomena that is independent of halo formation. This
series of papers will probe the limits of the influence of dark matter
on present-day galaxy properties, separating---and hopefully
simplifying---the problem of galaxy formation into those two regimes.

\vspace{1cm}
\noindent The authors wish to thank Andrew Hearin for useful
discussions. JLT acknowledges support from NSF grant AST-121189. ARW
was supported by a Moore Prize Fellowship through the Moore Center for
Theoretical Cosmology and Physics at Caltech and by a Carnegie
Fellowship in Theoretical Astrophysics at Carnegie Observatories. CC
acknowledges support from NASA grant NNX15AK14G, NSF grant
AST-1313280, and the Packard Foundation.


\appendix
\section{Supplementary halo age definitions}

In this appendix we show results from two more common halo age
proxies: the redshift at which half the halo mass forms, $\zhalf$, and
halo concentration, $\cvir$. $\zhalf$ is one of the most commonly used
halo age definitions in the field, while $\cvir$ is also widely used
as a proxy for halo age, given that halo clustering correlates well
with $\cvir$ at fixed $\mhalo$, and that $\cvir$ can be measured for
halos in a single snapshot, without having the create full halo
growth histories.

\begin{figure}
\psfig{file=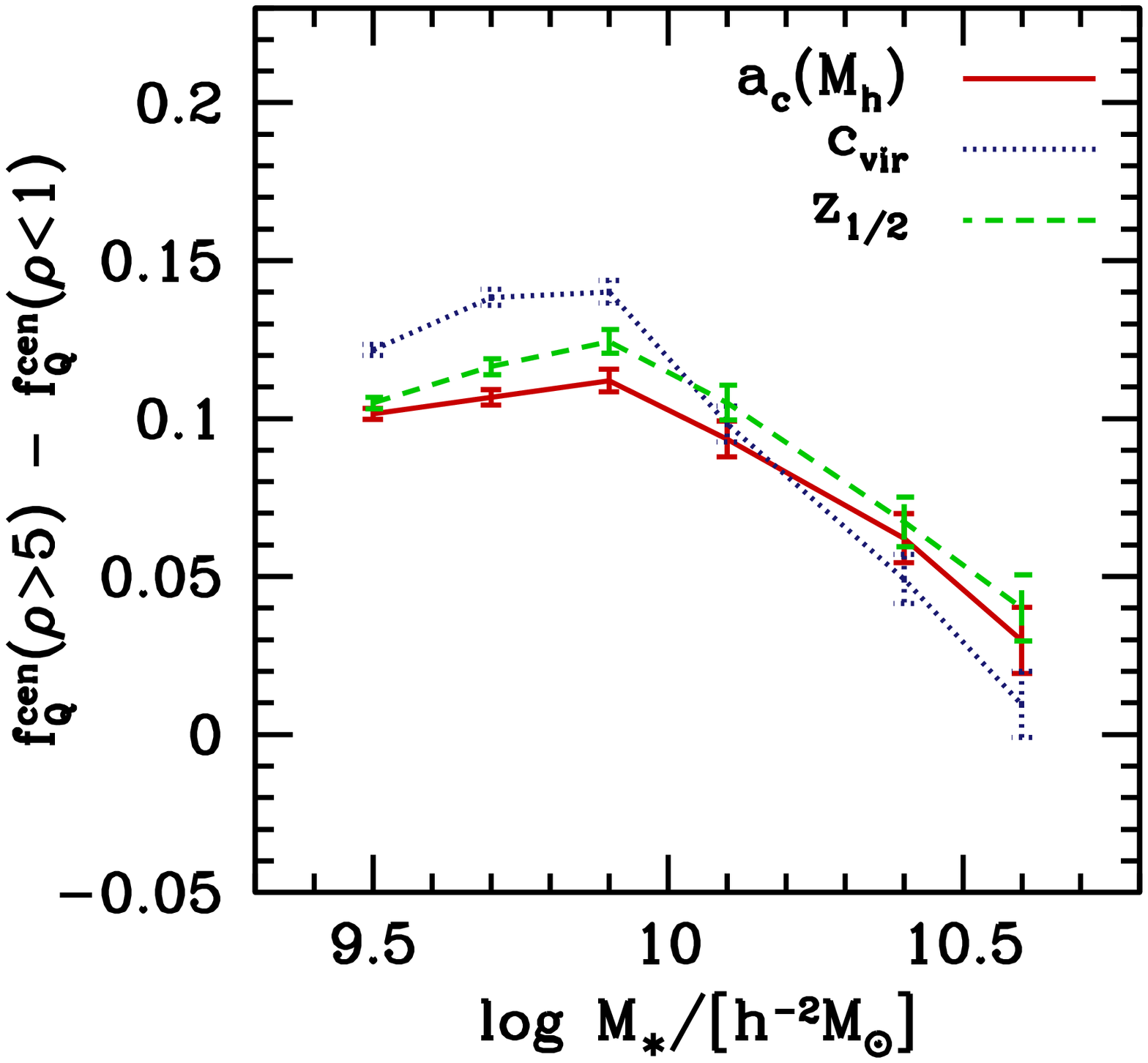,width=1.05\linewidth}
\caption{ \label{delta_fq_appendix} Same as Figure
  \ref{delta_fq}, but now showing the results for halo
  age-matching models in which halos are rank-ordered by $\zhalf$ and
  $\cvir$. For comparison, the model that uses $a_c(\mhalo)$ is also
  shown. }
\end{figure}

Figure \ref{delta_fq_appendix} reprises the assembly bias results
show in Figure \ref{delta_fq}, only now including the two new
halo age proxies listed about. The figure also shows the change in
$\fqcen$ for the $a_c(\mhalo)$ age-matching model, for
reference. There are slight quantitative differences in the assembly
bias induced using these two halo age proxies, but the overall results
are in good agreement with the fiducial age definitions used in the
main text.

\begin{figure}
\hspace{-1.5cm}
\psfig{file=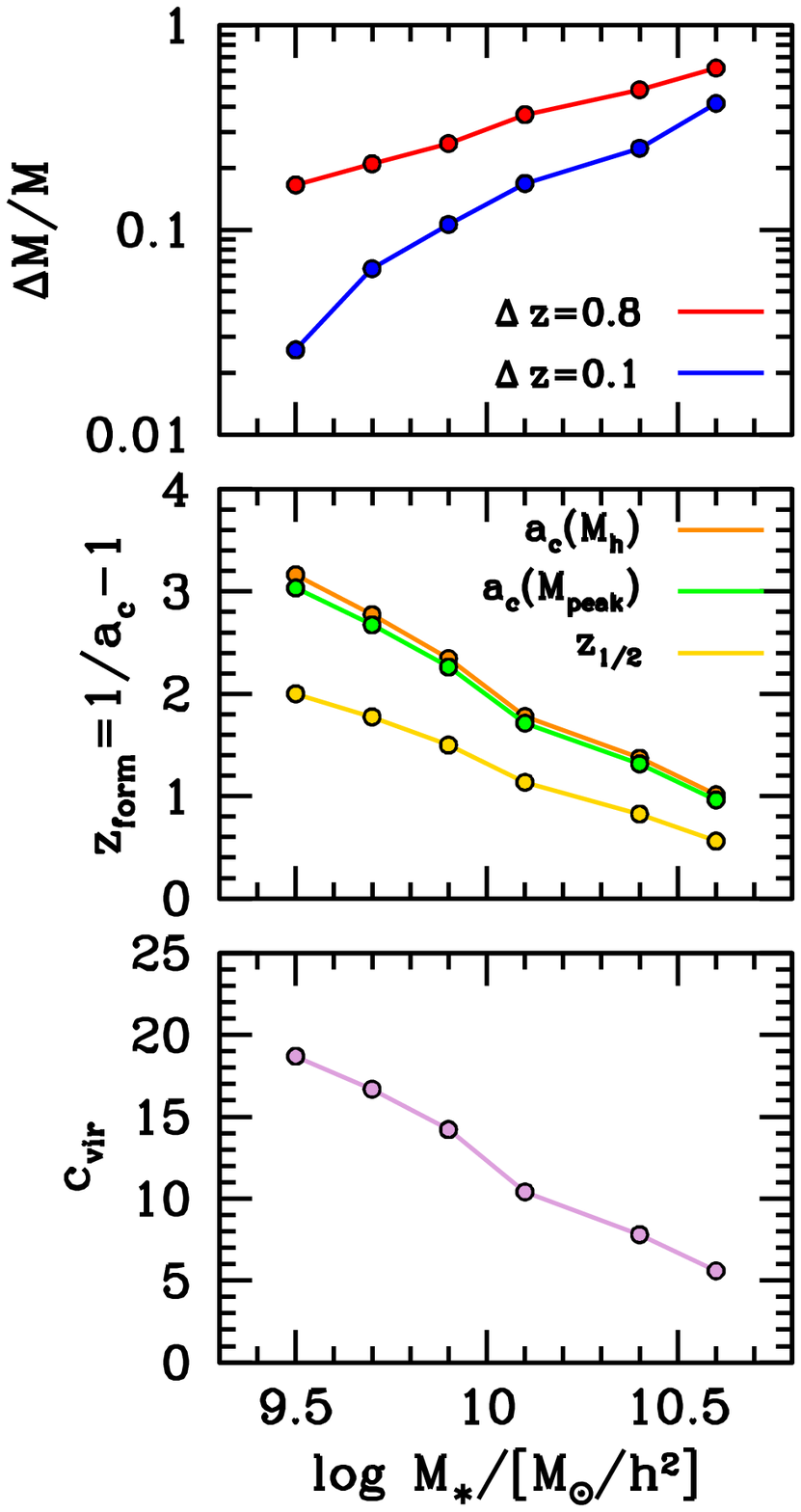,width=2.05\linewidth}
\caption{ \label{age_breaks} Each panel shows the break point between
  halos categorized as `old' and `young' for each halo age proxy. For
  each galaxy stellar mass, the old fraction is set to match the
  observed quenched fraction for central galaxies. {\it Top Panel:}
  Break point dividing old and young halos as determined by their
  fractional growth. Results are shown for redshift baselines $\Delta
  z=0.8$ and $\Delta z=0.1$. The trend of rising growth rate with
  $\log\mgal$ is due in small part to changing growth rates with halo
  mass, but mostly due to the rising $\fqcen$ with $\log\mgal$. {\it
    Middle Panel:} Break point using three different definitions of
  formation epoch as described in the text. {\it Bottom Panel:} Break
  point between old and young halos using $\cvir$ as the age proxy. }
\end{figure}

Figure \ref{age_breaks} shows the `break point' between halos being
classified as `old' and `young' for each of our halo age proxies. The
use of quotes around these terms is meant to stress that there is
possibly no physical significance to these values---they are somewhat
arbitrary dividing lines in continous distributions of halo
properties. But it is of interest to document the values required to
match the observed values of $\fqcen$ as a function of $\mgal$. The
top panel shows the break point when using fractional growth as our
age proxy; i.e., halos of $\mhalo\sim 10^{11.4}$ \hmsol\ (which house
galaxies of $mgal\sim 10^{9.5}$ $\hhmsol$), contain quenched galaxies
if their fractional growth rate is less than 17\%, when measured from
$z=0.8\rightarrow 0$. The middle panel shows the results using our
halo formation epoch estimates: $a_c(\mhalo)$, $a_c(\mpeak)$, and
$\zhalf$. We note that the formation epoch for halos using $\zhalf$ is
much smaller than when using $a_c$, but the amplitude and nature of
the assembly bias is nearly the same. In the bottom panel, we show the
values of $\cvir$ that delineate old from young halos. 

We note that the values shown for the $a_c$ model in Figure
\ref{age_breaks} are significantly smaller than those in the published
version of \cite{hearin_watson:13} (their Figure 2). An updated
version of their Figure 2, which will be submitted as an erratum, are
in good agreement with our results (A. Hearin, private
communication). We also note that the absolute values of the break
points do not alter the rank-ordering of the halos, and the results of
their paper are unchanged.

\bibliography{../risa}

\label{lastpage}

\end{document}

%% file: newcommands.tex
\def\hmpc{$h^{-1}$Mpc}

\def\msol{M$_\odot$}
\def\hmsol{$h^{-1}$M$_\odot$}

\def\rvir{R_{\rm vir}}

\def\om{\Omega_m}

\def\omb{\Omega_b}
\def\s8{\sigma_8}

\def\lcdm{$\Lambda$CDM}

\def\x2{$\chi^2$}
\def\hmsol{$h^{-1}\,$M$_\odot$}

\def\NNm1{\langle N(N-1) \rangle}

\def\m_star{M_\ast}

\def\lcdm{$\Lambda$CDM}

\def\om{\Omega_m}

\def\omb{\Omega_b}
\def\s8{\sigma_8}

\def\hmpc{$h^{-1}\,$Mpc}

\def\x2{$\chi^2$}
\def\hmsol{$h^{-1}\,$M$_\odot$}

\def\NNm1{\langle N(N-1) \rangle}

\def\p0{P_0(r)}